\documentclass[12pt]{article}
\usepackage{amssymb}
\usepackage{amsmath}
\usepackage{graphicx}

\begin{document}

\title{Unconventional state \\with two coexisting long-range orders \\for
frustrated Heisenberg model \\at quantum phase transition}
\author{A.V.\thinspace Mikheyenkov$^{\ast }$, N.A.\thinspace Kozlov,
A.F.\thinspace Barabanov, \\
%EndAName
Institute for High Pressure Physics RAS, \\ 142190, Troitsk,
Moscow region, Russia; \\Moscow Institute of Physics and
Technology, \\ 141700, 9, Institutskii
per., Dolgoprudny, Moscow Region, Russia\\
$^{\ast }${\normalsize e-mail: mikheen@bk.ru}}
\date{\today}
\maketitle

\begin{abstract}
For the frustrated two-dimensional $S=1/2$ antiferromagnetic Heisenberg
model close to quantum phase transition we consider the singlet ground
states retaining both translational and SU(2) symmetry. Besides usually
discussed checkerboard, spin-liquid and stripe states an unconventional
state with two coexisting long-range orders appears to be possible at
sufficiently large damping of spin excitations. The problem is treated in
the frames of self-consistent spherically symmetric approach.

PACS{74.72.-h, 71.27.+a, 75.20.-g}
\end{abstract}

\textbf{1.} The general quantum phase transition problem \cite{QPT-Rev1,
QPT-Rev2, QPT-Rev3} is commonly considered in the frames of 2D Heisenberg
quantum antiferromagnet. The main interest in this context is concentrated
around the maximally frustrated point $J_{2}/J_{1}\sim 0.5$ (frustration
parameter $p=J_{2}/(J_{1}+J_{2})\sim 0.3)$. In this region several
energetically competitive states with different symmetry appear to be
possible ground state $\Psi _{0}$\ at $T=0$. That is why the recent
theoretical efforts are aimed at quantum phase transition near $p\sim 0.3$
with several possible scenarios.

There is firstly a class of states which brake either translational $%
\widehat{T}_{\mathbf{l}}$ ($\mathbf{l}$ -- lattice vector) symmetry or spin
SU(2) symmetry of the Hamiltonian. These are semiclassical Neel checkerboard
state ($p<0.3$) and semiclassical stripe phase ($p>0.3$), where both $%
\widehat{T}_{\mathbf{l}}$ and SU(2) symmetries are broken. Another states of
the class are singlet $\langle S_{\mathbf{r}}^{z}\rangle =0$\ valence bond
crystal (VBC) states ($p\simeq 0.3$, $\widehat{T}_{\mathbf{l}}$ symmetry is
broken, SU(2) symmetry is restored). Spins in VBC are coupled in pairs
forming singlet valence bonds, the latter being arranged in a periodic
pattern. Usually only the simplest -- columnar and plaquette -- VBC are
considered. The aforenamed states are sketched in Fig.1.

Let us note, that an accurate comparison of competitive states is of course
permissible only within one and the same method.

Hereinafter we will consider only another class of pretender states --
singlet states which differ from semiclassical ones and do not brake neither
$\widehat{T}_{\mathbf{l}}$ nor SU(2) symmetry. The simplest state of this
class can be constructed in the frames of the mentioned nearest neighbour
valence bond\ approach. The wavefunction of this state is a sum of terms
represented in Fig.2a. Each term is given by disorderly placed valence
bonds, but the whole spin liquid (SL) state picture is periodical.

One can generalize the nearest neighbour valence bond concept and consider
also arbitrary range valence bonds (see Fig.2b). In the latter picture
spin-spin correlation functions $\langle S_{\mathbf{r}}^{z}|S_{\mathbf{0}%
}^{z}\rangle _{r\rightarrow \infty }$ for infinitely distant sites can
become non-zero under certain circumstances, thus opening the way to
consider the transition between spin liquid singlet state without long-range
order (LRO) to singlet state with LRO.

In the present work we investigate singlet states with and without LRO in
the frames of Kondo--Yamaji--Shimahara--Takada self-consistent spherically
symmetric theory (SST) \cite{Kondo72, Shimahara91}, adapted for frustrated
Heisenberg model \cite{BarBer94-both, We-Review}. SST does not brake neither
translational nor SU(2) symmetry. The method describes the singlet ground
state in terms of spin-spin correlation functions and allows to find the
triplet excitations spectrum. Note, that the arising description of Neel and
stripe states differs from common semiclassical ones, presented in Fig.1,
namely, a mean site spin in SST is equal to zero $\langle S_{\mathbf{r}%
}^{z}\rangle =0$ in any phase, sublattices are absent and the long-range
order (or its absence) is controlled by spin-spin correlators $\langle S_{%
\mathbf{r}}^{z}|S_{\mathbf{0}}^{z}\rangle _{r\rightarrow \infty }$ for
infinitely distant sites, or equivalently by the presence or absence of the
gap in the spin spectrum at the symmetrical points $(\pm \pi ,\pm \pi )$, $%
(0,\pm \pi )$, $(\pm \pi ,0)$.

In more detail, at zero temperature the structure factor $c_{\mathbf{q}%
}=\left\langle S_{\mathbf{q}}^{z}S_{\mathbf{-q}}^{z}\right\rangle $ in SST
is a superposition of smooth part $c_{0\mathbf{q}}$ and, for long-range
order cases, $\delta $-terms $c_{0}^{\nu }\delta (\mathbf{q}-\mathbf{q}%
_{0}^{\nu })$\ at symmetric points $\mathbf{q}_{0}^{\nu }$ of the Brilloine
zone. This means that in the coordinate space the spin-spin correlation
function acquires the form

\begin{equation}
c_{\mathbf{r}}=\langle S_{\mathbf{n+r}}^{z}|S_{\mathbf{n}}^{z}\rangle =\frac{%
1}{N}\sum_{\mathbf{q}}c_{0\mathbf{q}}e^{i\mathbf{qr}}+\sum_{\nu }c_{0}^{\nu
}e^{i\mathbf{q}_{0}^{\upsilon }\mathbf{r}}  \label{Cr0}
\end{equation}

The first term in the right-hand side of (\ref{Cr0}) defines local
correlations and it vanishes at infinite distance $r\rightarrow \infty $,
the second (condensation) part controls LRO. The existence of non-zero
condensation part is equivalent to zero spin gaps at the appropriate points $%
\mathbf{q}_{0}^{\upsilon }$\ of the Brillouin zone.

In particular for spherically symmetric Neel-type state (SST-Neel) $\mathbf{q%
}_{0}^{\nu =1}=\mathbf{Q}=(\pi ,\pi )$ and the long-range order has a
checkerboard\ pattern

\begin{equation}
\langle S_{\mathbf{r}}^{z}|S_{\mathbf{0}}^{z}\rangle _{|\mathbf{r|}%
\rightarrow \infty }=c_{0}^{Neel}(-1)^{n_{x}+n_{y}}  \label{Cond_Neel}
\end{equation}

where $\mathbf{r}=n_{x}\mathbf{g}_{x}+n_{y}\mathbf{g}_{y}$, and $\mathbf{g}%
_{x}=(1,0)$, $\mathbf{g}_{y}=(0,1)$ are lattice vectors. The effective
magnetization $m^{Neel}=\sqrt{c_{0}^{Neel}}$depends on frustration. SST-Neel
state is realized at low frustration. $m^{Neel}$ decreases from its maximal
value at $p=0$ to zero at point $p_{N}^{\ast }$, where LRO disappears ($%
p_{N}^{\ast }\sim 0.2$, \cite{We09, Schmalfuss06}). Let us remind,
that in SST $\langle S_{\mathbf{r}}^{z}\rangle =0$ in any phase.

Spherical symmetric stripe state (SST-Stripe) differs from the
presented in Fig.1 semiclassical picture also in the following
sense -- it is the coherent superposition of stripe pictures with
horizontal (as in Fig.1) and vertical stripes. For this state
$\mathbf{q}_{0}^{\nu =2,3}=\mathbf{X} =(0,\pi )$, $(\pi ,0)$ and
the long-range order pattern is

\begin{equation}
\langle S_{\mathbf{r}}^{z}|S_{\mathbf{0}}^{z}\rangle _{|\mathbf{r|}%
\rightarrow \infty }=\frac{c_{0}^{Stripe}}{2}[(-1)^{n_{x}}+(-1)^{n_{y}}]
\label{Cond_Stripe}
\end{equation}

SST-Stripe state is realized for higher frustrations. The effective
magnetization $m^{Stripe}=\sqrt{c_{0}^{Stripe}}$ increases from zero at $%
p_{S}^{\ast }\sim 0.5$ \cite{We-Review}\ to the maximal value at $p=1$.

The solution with $c_{0}^{Neel}=c_{0}^{Stripe}=0$, i.e. without
long-range order, corresponds to SL. In this case spin gaps in the
symmetrical points of the Brilloine zone are opened, their value
depending on frustration. The correlation length is defined by the
lowest spin gap and also depends on frustratrion. In this sense
the solution involved is the arbitrary range spin liquid (compare
Fig.2b). Self-consistent SL solution is realized in the
intermediate frustration region $p\sim 0.3$ \cite{We-Review, We09,
Schmalfuss06}.

Note, that besides three aforenamed types of solutions the approach allows
to consider the exotic state with two mutually penetrating long-range orders
-- checkerboard and stripe. The spin gaps in this state must be closed both
at $\mathbf{Q}$ and at $\mathbf{X}$. The main result of \ this work is that
at sufficiently large damping of spin excitations the solution of this type
does exist and at the point of quantum phase transition is degenerate with
another (two) solutions.

The paper is organized as follows. In part 2 we briefly remind the SST
calculation procedure in the mean field realization. In part 3 the damping
of spin excitations is introduced in the simplest semiphenomenological way.
As it is shown in part 4, at sufficiently large but still realistic damping
there appears a new type of self-consistent solution -- with two long-range
orders.\ In passing\ it is demonstrated that the width of spin liquid
solution region strongly depends on damping. This part is also devoted to
brief discussion.

\textbf{2.} The Hamiltonian of the frustrated Heisenberg model for $S=1/2$
spins on a square lattice is

\begin{equation}
\widehat{H}=\frac{1}{2}J_{1}\sum_{\mathbf{i,g}}\overrightarrow{\mathbf{S}}_{%
\mathbf{i}}\overrightarrow{\mathbf{S}}_{\mathbf{i+g}}+\frac{1}{2}J_{2}\sum_{%
\mathbf{i,d}}\overrightarrow{\mathbf{S}}_{\mathbf{i}}\overrightarrow{\mathbf{%
S}}_{\mathbf{i+d}}  \label{Hamilt}
\end{equation}

here the first term in the right-hand side stands for the nearest-neighbour
interaction ($J_{1}$ -- antiferromagnetic exchange constant, $\mathbf{g}$ --
nearest-neighbour vector), and the second -- for the next-nearest-neighbour
one (with corresponding quantities $J_{2}$ and $\mathbf{d}$). Standard
parameter $p$ is a measure of frustration $p=J_{2}/J$, $J_{1}=(1-p)J$, $%
J_{2}=pJ$, $J=J_{1}+J_{2}$. All the energetic parameters are expressed in
the units of $J$, hereafter\ we put $J=1$.

In the frames of SST the calculations come to the chain of equations of
motion for spin retarded Green's function $G_{\mathbf{nm}}^{z}=\langle S_{%
\mathbf{n}}^{z}|S_{\mathbf{m}}^{z}\rangle _{\omega +i\delta
}=-i\int\limits_{0}^{\infty }dt\,e^{i\omega t}\langle \lbrack S_{\mathbf{i}%
}^{z}(t),S_{\mathbf{j}}^{z}]\rangle $. The chain is closed at the second
step by the approximation of the following type

\begin{equation}
\begin{array}{c}
S_{\mathbf{n+g}_{1}+\mathbf{g}_{2}}^{j}S_{\mathbf{n+g}_{1}}^{l}S_{\mathbf{n}%
}^{\gamma }\approx \alpha _{\mathbf{g}}(\delta _{jl}\left\langle S_{\mathbf{%
n+g}_{1}+\mathbf{g}_{2}}^{j}S_{\mathbf{n+g}_{1}}^{l}\right\rangle S_{\mathbf{%
n}}^{\gamma }+ \\
+\delta _{l\gamma }\left\langle S_{\mathbf{n+g}_{1}}^{l}S_{\mathbf{n}%
}^{\gamma }\right\rangle S_{\mathbf{n+g}_{1}+\mathbf{g}_{2}}^{j})+\alpha _{%
\mathbf{g}_{1}+\mathbf{g}_{2}}\delta _{j\gamma }\left\langle S_{\mathbf{n+g}%
_{1}+\mathbf{g}_{2}}^{j}S_{\mathbf{n}}^{\gamma }\right\rangle S_{\mathbf{n+g}%
_{1}}^{l}%
\end{array}
\label{decouple}
\end{equation}

where $\alpha _{\mathbf{r}}$ -- so called vertex corrections \cite%
{Shimahara91}. In the mean-field approximation Green's function $G^{z}(%
\mathbf{q},\omega )=\langle S_{\mathbf{q}}^{z}|S_{-\mathbf{q}}^{z}\rangle
_{\omega }$ has the form

\begin{equation}
G^{z}(\mathbf{q},\omega )=\frac{F_{\mathbf{q}}}{\omega ^{2}-\omega _{\mathbf{%
q}}^{2}},\quad S_{\mathbf{q}}^{z}=\frac{1}{\sqrt{N}}{\,}\sum\limits_{\mathbf{%
q}}e^{-i\mathbf{qr}}S_{\mathbf{r}}^{z}  \label{Gq}
\end{equation}

where the numerator $F_{\mathbf{q}}$ and the spin excitations spectrum $%
\omega _{\mathbf{q}}$ are functions of lattice sums, involving spin--spin
pair correlation functions $c_{\mathbf{r}}=\langle S_{\mathbf{n+r}}^{z}|S_{%
\mathbf{n}}^{z}\rangle $ for first five coordination spheres (see
\cite {We09, We06-07} for details).

$G^{z}(\mathbf{q},\omega )$ defines the structure factor, i.e.
Fourier-transform of a correlation function

\begin{equation}
c_{\mathbf{q}}=\left\langle S_{\mathbf{q}}^{z}S_{\mathbf{-q}%
}^{z}\right\rangle =-\frac{1}{\pi }\int d\omega \,m(\omega )\,\mathbf{Im}%
\,G^{z}(\mathbf{q},\omega )=\frac{F_{\mathbf{q}}}{2\omega _{\mathbf{q}}}%
\left( 2m(\omega _{\mathbf{q}})+1\right) ;  \label{Cq}
\end{equation}%
\begin{equation*}
m(\omega )=\left( e^{\frac{\omega }{T}}-1\right) ^{-1}
\end{equation*}

which, in turn, leads to five self-consistent equations:

\begin{equation}
c_{\mathbf{r}}=\frac{1}{N}\sum_{\mathbf{q}}c_{\mathbf{q}}e^{i\mathbf{qr}%
};\quad (\mathbf{r}=\mathbf{g},\mathbf{d},2\mathbf{g},2\mathbf{d},\mathbf{g}+%
\mathbf{d})  \label{Cr2}
\end{equation}

The system of equations (\ref{Cr2}) is then solved numerically.

At $T=0$ the structure factor $c_{\mathbf{q}}$ (\ref{Cq}) is a superposition
of smooth part and, possibly, $\delta $-terms $c_{0}^{\nu }\delta (\mathbf{q}%
-\mathbf{q}_{0}^{\nu })$\ at symmetric points $\mathbf{q}_{0}^{\nu }$ of the
Brilloine zone. The latter solution corresponds to the presence of
long-range order, the condensation part appears in Eq.(\ref{Cr2}), and (\ref%
{Cr2}) transforms into

\begin{equation}
c_{\mathbf{r}}=\frac{1}{N}\sum_{\mathbf{q}}c_{0\mathbf{q}}e^{i\mathbf{qr}%
}+\sum_{\nu }c_{0}^{\nu }e^{i\mathbf{q}_{0}^{\upsilon }\mathbf{r}}
\label{Cr-cond}
\end{equation}

where $c_{0\mathbf{q}}=\frac{F_{\mathbf{q}}}{2\omega _{\mathbf{q}}}$\
defines local correlations, and $c_{0}^{\nu }$ controls the long-range
order. Non-zero condensation part leads to additional condition in the
self-consistent system of equations (\ref{Cr2})\ --- the zero spin gap at
the appropriate point of the Brillouin zone ($\mathbf{q}_{0}^{\nu =1}=%
\mathbf{Q}=(\pi ,\pi )$ for SST-Neel and $\mathbf{q}_{0}^{\nu =2,3}=\mathbf{X%
}=(0,\pi )$, $(\pi ,0)$ for SST-Stripe phase). Solution with all $c_{0}^{\nu
}=0$, i.e. without long-range order corresponds to spin liquid. As it is
seen from (\ref{Cr-cond}), in particular, the scheme allows to search for
the states with several mutually penetrating long-range orders, defined by
the points $\mathbf{q}_{0}^{\nu }$.

\textbf{3.} The described starting approach leads to spin excitations
without damping, being the mean-field realization of SST. Nevertheless, at
this stage it is possible to trace the evolution of the ground state with
the frustration increase from SST-Neel phase to spin liquid and further to
SST-Stripe phase \cite{We-Review, We06-07}.

But, as it was shown in \cite{ We09, We06-07}, the damping of spin
excitations is qualitatively important, even if it is introduced
in the semiphenomenological way. The scheme, based on formally
exact expression for the Green's function \cite{BarMaks95}, in the
simplest case is reduced to the following Green's function form
(instead of\ (\ref{Gq})):

\begin{equation}
G^{z}(\mathbf{q},\omega )=\frac{F_{\mathbf{q}}}{\omega ^{2}-\omega _{\mathbf{%
q}}^{2}+i\omega \gamma }  \label{hi_our}
\end{equation}

here only the imaginary part of the polarization operator is taken into
account and it is written down as the trivial odd function of $\omega $.
Damping constant $\gamma $ must be considered as an external parameter,
defined by whatever additional arguments.

The rest of the self-consistent calculations procedure remains the same (of
course, it must be recalculated at any fixed $\gamma $).

\textbf{4.} As it was mentioned above, in principle one can seek four types
of self-consistent solutions. The first is the spin-liquid solution without
long-range order $c_{|\mathbf{r|}\rightarrow \infty }\rightarrow 0$. There
are also two types of the "sole" long-range order solutions: with
checkerboard $c_{|\mathbf{r|}\rightarrow \infty }\sim e^{i\mathbf{Qr}}\sim
(-1)^{n_{x}+n_{y}}$ and quantum stripe $c_{|\mathbf{r|}\rightarrow \infty
}\sim e^{i\mathbf{Xr}}\sim \frac{1}{2}[(-1)^{n_{x}}+(-1)^{n_{y}}]$
correlators pattern. The fourth possible type of solution corresponds to two
mutually penetrating long-range orders, when the spin-spin correlator at
infinitely distant sites is the linear combination on checkerboard and
stripe laws.

In the mean-field approximation with zero damping only three self-consistent
solutions are realized in SST -- SST-Neel for small frustration $p$,
spin-liquid for intermediate $p$ and SST-Stripe for large frustration \cite%
{We-Review}.

Fig.3 represents the energy for all types of self-consistent solutions
(found in the present work) in the whole range of frustration parameter $p$
for different values of damping parameter $\gamma $. It is seen, that at
small $\gamma $ $(\gamma \lesssim 0.3)$\ the picture remains qualitatively
unchanged: for different values of $\gamma $ only three types of solutions
are realizes.

With growing damping $\gamma $ the points of transitions $p_{N}^{\ast }$
(Neel$\longleftrightarrow $Liquid) and $p_{S}^{\ast }$ (Liquid$%
\longleftrightarrow $Stripe) are shifted towards each other. It was shown in
our previous work \cite{We09}, that $p$-region of SL solution shrinks from
the left side with the increase of $\gamma $, i.e. the value $p_{N}^{\ast }$
increases. Current results show, that the same is true for the right side of
SL region, corresponding to SL$\longleftrightarrow $Stripe transition.

But at $\gamma \gtrsim 0.3$ for intermediate frustrations $p\approx 0.3$
there appears another self-consistent solution -- with two coexisting
long-range orders. For $0.3\lesssim \gamma \lesssim 0.6$ this solution is
metastable in the whole region of its existence and is disconnected by
energy with the other solutions.

At $\gamma \gtrsim 0.6$ this biordered solution is energetically connected
with the SL one for $p\lesssim 0.3$ and with stripe state for $p\gtrsim 0.3$%
, in the intermediate interval it remains metastable. This is clearly seen
from Fig.4, where for $\gamma =0.65$ the energies of all four solutions --
SST-Neel, SST-Stripe, SL and biordered are depicted. The corresponding
condensates for long-range ordered solutions are also shown.

Note that the value $\gamma \sim 0.6$ can be considered as absolutely
realistic \cite{Prelovsek03-gam1, Sherman03-gam2}.

As it is seen from Figs.3--4, three states of different types
(SST-Neel, SL and biordered) at $\gamma \gtrsim 0.5$\ are
degenerate at the quantum transition point $p^{QPT}\approx 0.28$.
Thus we come to a conclusion, that in the maximally frustrated
region quantum fluctuations in addition to usually discussed
states (checkerboard, SL and other mentioned above states) can
select the exotic state with two long-range orders.

Let us mention, that in the biordered state at $p\gtrsim p^{QPT}$ with
decrease of $p$ the checkerboard condensate increases and the stripe
condensate rapidly decreases.

The typical spin excitations spectrum $\omega _{\mathbf{q}}$\ for
this state is presented in Fig.5. The spectrum is gapless at
points $\mathbf{X}$ and $ \mathbf{Q}$.

As it was noted above for intermediate damping $0.3\lesssim \gamma \lesssim
0.6$ the biordered solution is disconnected by energy with other ones (see
Fig.3, curves for $\gamma =0.35$ and $\gamma =0.45$). To reveal the states
which could be energetically connected with the biordered state, one must
examine other candidates, for example the helicoidal solution with $\mathbf{q%
}_{0}^{\nu }=(q,\pi )$, $(\pi ,q)$. Its semiclassical prototype is the
helicoidal state, which is realized in $J_{1}-J_{2}-J_{3}$ Heisenberg model
\cite{Mambrini06}. Nevertheless this speculation is a matter of further
study.

In conclusion, we have found that in the frames of self-consistent
spherically symmetric approach for the 2D frustrated Heisenberg model at $T=0
$ the unconventional state with two coexisting long-range orders can appear
near the point of quantum phase transition $p^{QPT}\approx 0.28$ at
sufficiently large damping of spin excitations.

The work was supported by the Russian Fund of Fundamental Investigations.

\newpage
\begin{figure}
    \includegraphics[width=10cm]{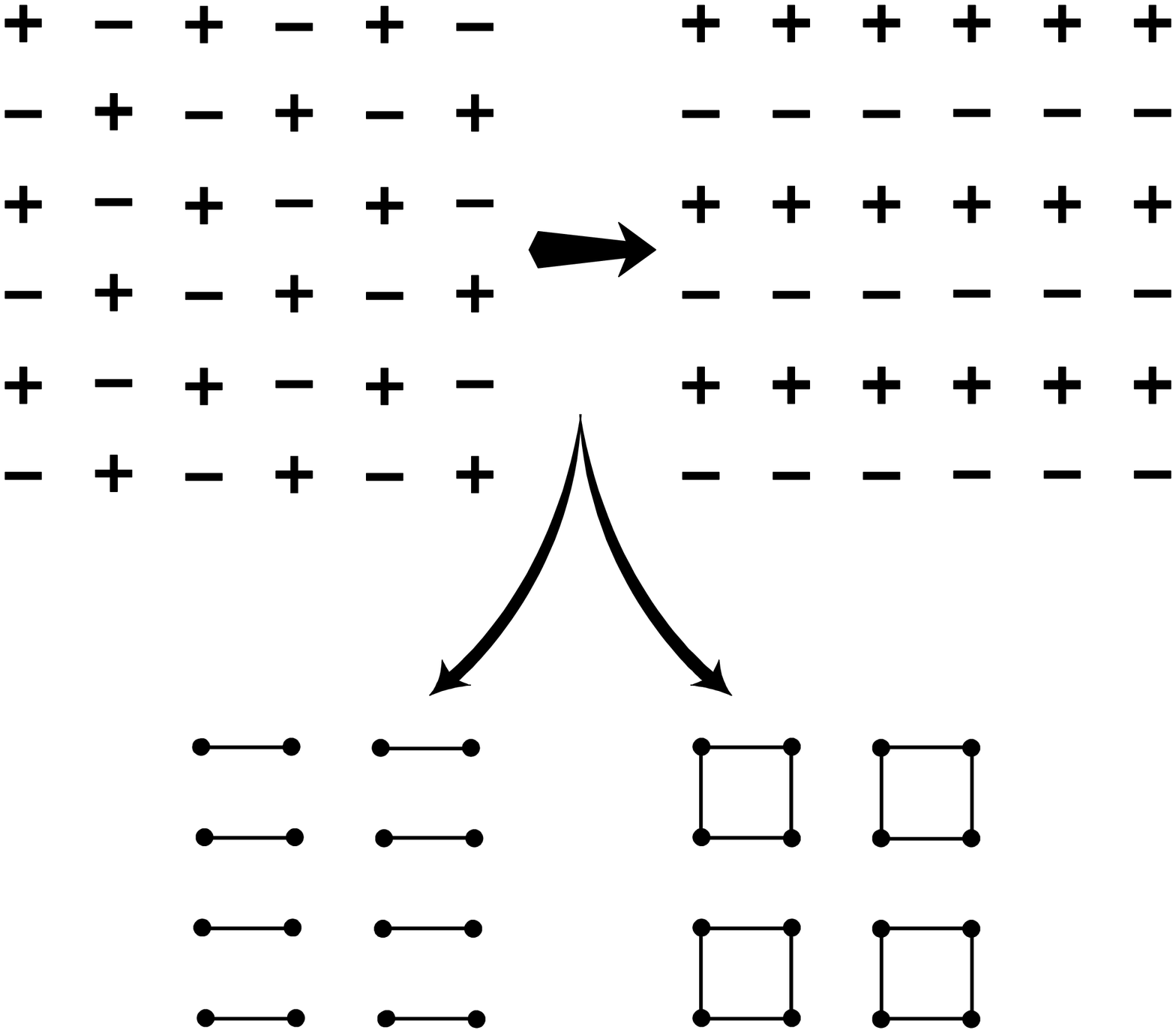}\\
  \caption{Commonly considered states for 2D Heisenberg model.
  Upper left -- semiclassical Neel checkerboard state,
  upper right -- semiclassical stripe state, thick arrow
  implies the transition with the frustration increase.
  Lower left -- columnar VBC, lower right -- plaquette VBC}
\end{figure}

\begin{figure}
    \includegraphics[width=10cm]{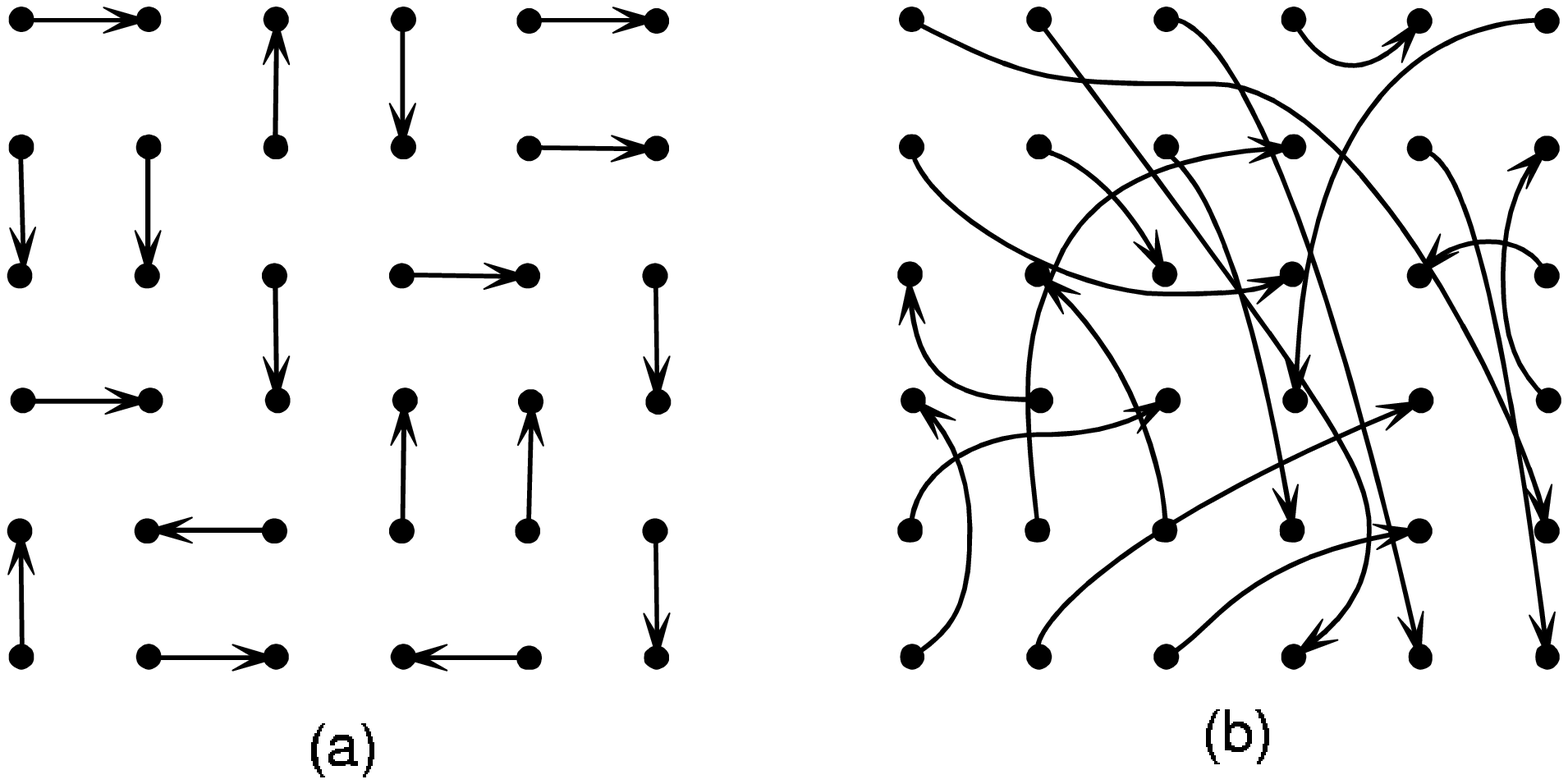}\\
  \caption{a) Nearest neighbour valence bond spin-liquid state.
b) Arbitrary range valence bond spin-liquid state. The oriented
bond between two sites $i$ and $j$ stands for singlet valence bond
$(1/\sqrt{2})(|\uparrow _{i}\downarrow _{j}\rangle -|\downarrow
_{i}\uparrow _{j}\rangle )$.}
\end{figure}

\newpage
\begin{figure}
    \includegraphics[width=14cm]{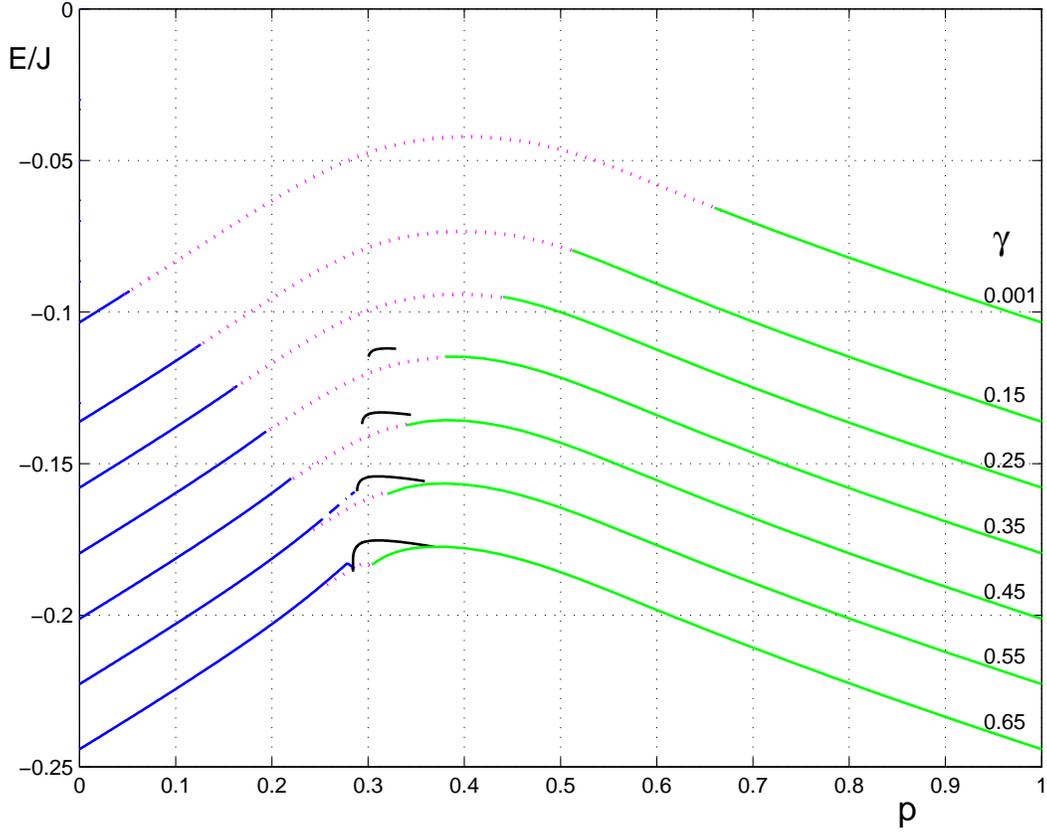}\\
  \caption{The energy for all found types of self-consistent solutions
  for $0\le p \le 1$ and different values of damping $\gamma$.
  Solid lines on the left correspond to SST-Neel solutions,
  solid lines on the right -- SST-Stripe, intermediate dotted
  lines are SL solutions, small solid curves above SL at $p\sim 0.3$
  correspond to biordered solution.
  The ordinate axis scale fits the curves with $\gamma=0.001$,
  other lines are downshifted by $0.2\gamma$.}
\end{figure}

\begin{figure}
    \includegraphics[width=14cm]{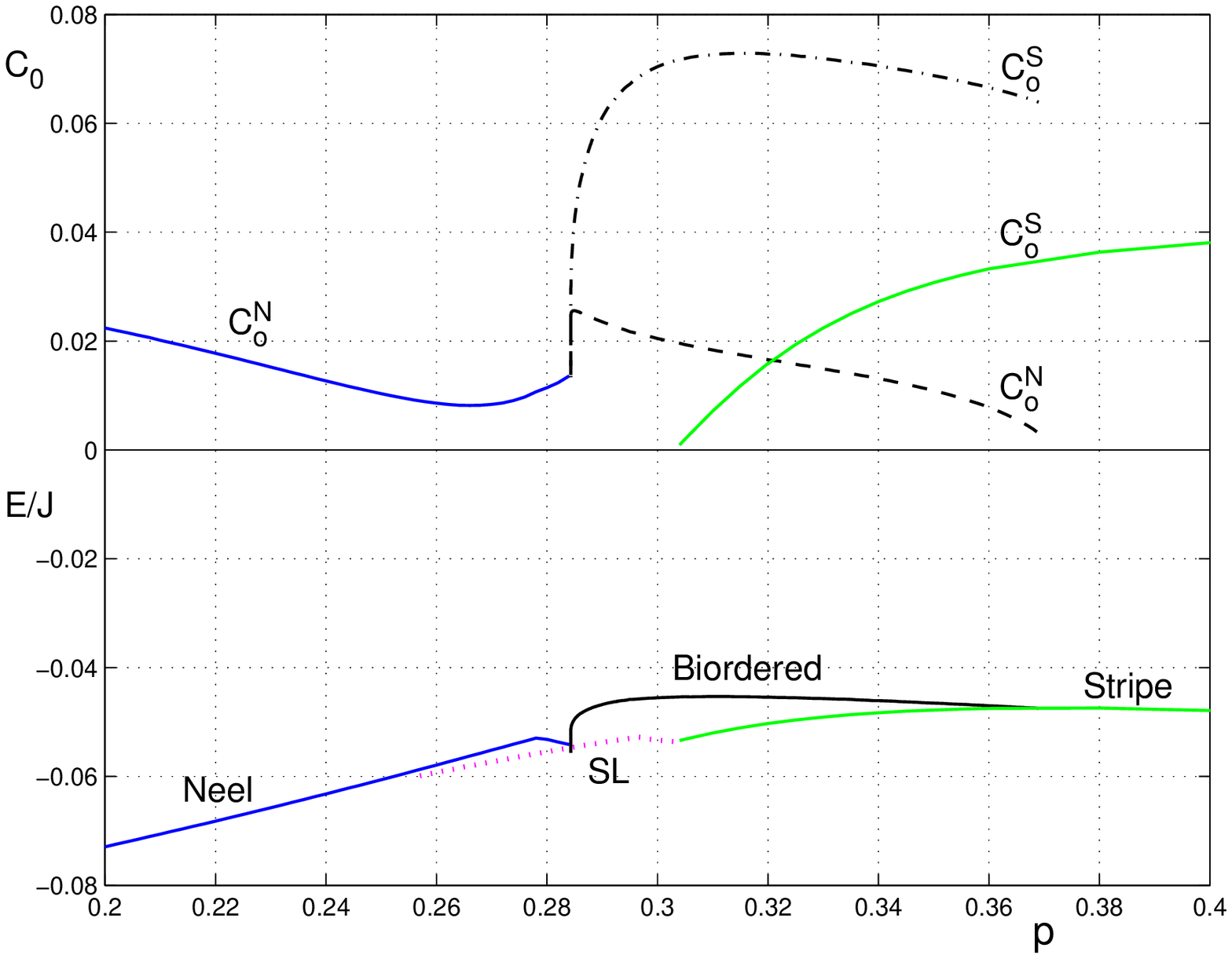}\\
  \caption{The energy of SST-Neel, SL, biordered and SST-Stripe
  states in the vicinity of quantum phase transition for damping
  $\gamma=0.65$.
  \newline The upper graphs show the condensates for the
  LRO solutions. For checkerboard LRO
  $c_{|\mathbf{r|}\rightarrow \infty }\sim C_0^N
  (-1)^{n_{x}+n_{y}}$,
  for stripe LRO $c_{|\mathbf{r|}\rightarrow \infty }\sim C_0^S
  (1/2)[(-1)^{n_{x}}+(-1)^{n_{y}}]$.
  Solid lines are for the SST-Neel (left) and SST-Stripe (right)
  states. The dashed and dashed-dotted curves are checkerboard
  and stripe condensates for the biordered state.}
\end{figure}

\newpage
\begin{figure}
    \includegraphics[width=14cm]{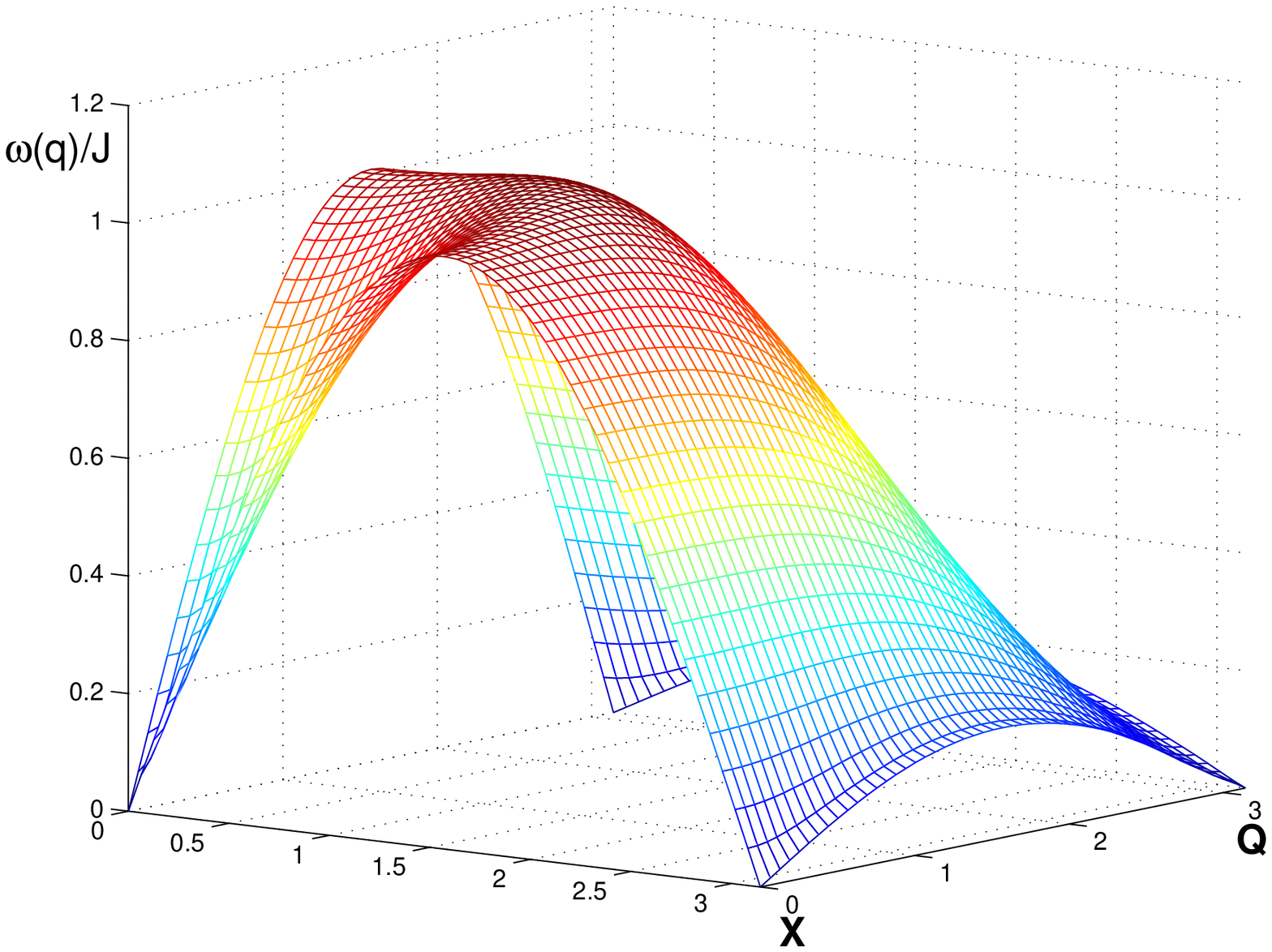}\\
  \caption{Spin excitations spectrum $\omega_{\mathbf{q}}$ for the state
  with two coexisting long-range orders. Frustration and damping
  parameters are $p=0.285$ and $\gamma=0.65$.
  The spectrum is gapless at points $\mathbf{X}$ and $\mathbf{Q}$}
\end{figure}


\begin{thebibliography}{99}
\bibitem{QPT-Rev1} S. Sachdev, Quantum Phase Transitions, Cambridge Univ.
Press, 1999

\bibitem{QPT-Rev2} S. Sachdev, Nature Physics, 4 (2008) 173, \
cond-mat/0711.3015v4

\bibitem{QPT-Rev3} P. Fulde, P. Thalmeier and G. Zwicknagl, Strongly
Correlated Electrons, in: Solid State Physics, Advances in Research and
Applications, Vol. 60, 2006, p.1, cond-mat/0607165 v1.

\bibitem{Kondo72} J. Kondo, K. Yamaji, Prog. Theor. Phys. 47 (1972) 807.

\bibitem{Shimahara91} H. Shimahara and S. Takada, J. Phys. Soc. Jpn. 60,
2394 (1991).

\bibitem{BarBer94-both} A.F. Barabanov, V.M. Berezovsky, JETP 79 (1994) 627;
A.F. Barabanov, V.M. Berezovsky, J. Phys. Soc. Jpn. 63 (1994) 3974.

\bibitem{We-Review} A.F. Barabanov, L.A. Maksimov, A.V. Mikheenkov, Spin
polaron in the cuprate superconductor: Interpretation of the ARPES results,
in: N.M. Plakida (Ed.), Spectroscopy of High-Tc Superconductors. A
Theoretical View, Taylor\&Francis, 2003, p. 1.

\bibitem{We09} A.V. Mikheyenkov, N.A. Kozlov, A.F. Barabanov, Phys. Lett. A
373 (2009) 693.

\bibitem{Schmalfuss06} D. Schmalfuss, et al., Phys. Rev. Lett. 97 (2006)
157201;

\bibitem{We06-07} A.V. Mikheyenkov, A.F. Barabanov, N.A. Kozlov, Phys. Lett.
A 354 (2006) 320; A.F. Barabanov, A.V. Mikheenkov, A.M. Belemuk, Phys. Lett.
A 365 (2007) 469; A.V. Mikheenkov, A.F. Barabanov, JETP 105 (2007) 347, Zh.
Eksp. Teor. Fiz. 132 (2007) 392.

\bibitem{BarMaks95} A.F. Barabanov, L.A. Maksimov, Phys. Lett. A 207 (1995)
390.

\bibitem{Prelovsek03-gam1} P. Prelov\v{s}ek, et al., Phys. Rev. Lett. 92
(2004) 027002.

\bibitem{Sherman03-gam2} A. Sherman, M. Schreiber, Phys. Rev. B 68 (2003)
094519.

\bibitem{Mambrini06} A. Moreo, et al., Phys. Rev. B 42 (1990) 6283;
M. Mambrini, et al., Phys. Rev. B 74 (2006) 144422.
\end{thebibliography}
\end{document}